# An IoT based Active Building Surveillance System using Raspberry Pi and NodeMCU

Sruthy. S, S. Yamuna, and Sudhish N. George, *Member, IEEE*

*Abstract*— Internet of Things (IoT) has emerged with a motive to automate the human life. It can be visualized as a network of connected things which is capable of providing intelligent services. This paper presents an IoT based security surveillance system in buildings using Raspberry Pi Single Board Computer (SBC) and NodeMCU (WiFi/IoT module). This system comprises of wireless sensor nodes and a controller section for surveillance. Intrusion detection with face detection and recognition, fire detection, remote user alerts, live video streaming and portability are the prime features of the system. The use of face recognition feature in intrusion detection makes the system more efficient by identifying the known and unknown person in restricted areas. WiFi module processes the sensor based events and sends the sensor status to controller section. Upon receiving the event notification, the controller enables the camera for capturing the event, alerts the user via email, phone call and Short Message Service (SMS) and places the live video of event on webpage. The use of WiFi module makes the node compact, cost effective and easy to use. The biggest advantage of the system is that the user can seek surveillance from anywhere in the world and can respond according to the situations.

## I. Introduction

Digitalization can be considered as a revolution of digital technologies where people and things are interconnected for every possible need. Its aim is to benefit the people with the help of internet services. The evolution of internet can be realized as a sequence of steps as follows: Initially we introduced the concept of network of computers known as Internet of Computers. When people got connected to internet through social media and other networking games, Internet of People emerged. Recently, a rapidly growing network of things has been launched and it is known as Internet of Things (IoT). The basic structure of IoT is shown in Fig. 1.

Sruthy. S is with the National Institute of Technology, Calicut, India (e-mail: sruthysukumaran1@gmail.com).

S. Yamuna is with the National Institute of Technology, Calicut, India (e-mail: yamusubru@gmail.com).

Sudhish. N. George is with the National Institute of Technology, Calicut, India (e-mail: sudhish@nitc.ac.in).

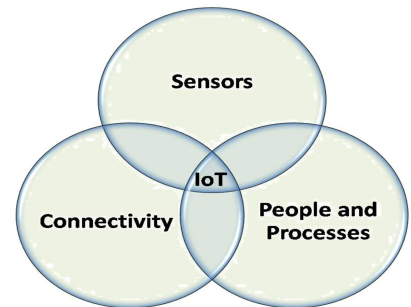

Fig. 1. Basic Structure of IoT

It consists of sensors, connectivity and people and processes. Things connected to internet collects data using embedded sensors. Then, the data is digitized and placed into networks. With networked data, we can create bidirectional systems where people and processes integrate for better decision making. The IoT based technologies minimizes human effort, improves resource utility and creates smart applications like security system, medical and healthcare systems, environmental monitoring, weather control, transportation etc.

As part of the digitization of India, a campaign known as 'Digital India' has been launched by the Prime Minister of India, to transform our country into a digitally empowered society. Its main motive is to connect rural areas with high speed internet networks. The digital space of our country is undergoing various transformations and the most recent entry into the digital space is IoT. As part of the Digital India Program, our government has taken several initiatives for the development of IoT industry in the nation. They have come out with a draft policy on IoT which mainly focuses on smart city developments such as smart parking, intelligent transport system, smart energy, citizen safety etc. The other notable initiatives of IoT are delivering education, health, security and financial services to remote areas. In this manner, IoT provides a bridge between rural and urban areas.

This paper details the design and implementation of an IoT based security system in home/building. In the present scenario, ensuring safety and security has become an inevitable essentiality. Since it is well known that influence of modern technology has reached its peak, demand for security systems are going up progressively. Modern home needs intelligent systems with minimum human effort. With the advent of digital and wireless technologies, automated security systems become more intelligent. Surveillance camera helps the user to get a remote view of his home and the sensor networks add extra security features depending on the type of sensors. Adding WiFi to security systems enables faster data transmission, and it will help the user to monitor and control the systems globally.

*A. Overview of Existing Systems*

Among the existing surveillance techniques, Closed Circuit Television Monitoring (CCTV) system is the most commonly used one. But it has its own limitations. It is a passive monitoring device and it needs continuous human intervention for monitoring. The investigation is a little bit hectic thing since all the previously recorded videos need to be watched manually. Moreover, files can be corrupted very easily and this technique is costly too. These limitations lead to the development of active surveillance system. Several researchers have come up with the idea of active surveillance systems in various papers. Most of the papers utilize the advantage of Wireless Sensor Networks (WSN) for surveillance. Since the sensor nodes being wireless, they can be placed anywhere inside the building, thus it achieves portability in deployment.

Zhao *et al.* proposed a WSN based surveillance system monitored by Programmable System on Chip (PSoC) devices [1]. Here Zigbee module is used for wireless transmission. The system mainly concentrates on sensor based alerts and it lacks improved techniques like camera, web server for uploading files etc.

Balasubramanian *et al.* presented the analysis of different remote control techniques used in home security systems [2]. It explains several schemes for remote control and monitoring of home appliances and security systems. It lacks modern surveillance techniques like web server, live streaming, email alert etc.

Bai *et al.* in their paper detailed the implementation of embedded home surveillance system using multiple ultrasonic sensors [3]. Intruder detection is the main feature of the system. It has Webcam to capture the image and web server to upload the images. It does not have modern alerting techniques such as email, SMS etc. Also, the development cost of the system is high due to the use of multiple sensors, amplifiers etc.

Song *et al.* in their paper discussed about automatic docking system for recharging home surveillance robots [4]. This system consists of a robot with automatic recharging capability for home security. Obstacle detection is the main feature of the system. Robot has a camera to capture the event and a dedicated computer to store the images. It does not contain user alerting techniques and it needs continuous monitoring. Also, the use of computer increases the power consumption.

Bai *et al.* proposed the design and implementation of a home embedded surveillance system with ultra-low alert power [5]. Here the system uses multiple sensor groups with low power consumption for the detection of an intruder. It lacks other sensor alerts, SMS alerts, live streaming and email alerts etc.

Rakesh *et al.* proposed an improved real time home security system using beagle board and Zigbee [6]. Remote alert on fire and intruder detection are the main features of the system. It uses improved techniques such as camera, Global System for Mobile (GSM), File Transfer Protocol (FTP) server etc. But it is not utilizing the advantage of live streaming and alerting techniques such as phone calls, email etc.

Ansari *et al.* proposed an Internet of Things Approach for Motion Detection using Raspberry Pi [7]. It utilizes FTP server for camera feeds and it alerts user through email. The system does not have SMS and phone call alerts and other sensor alerts such as detection of fire, gas etc.

Muheden *et al.* in their paper explained a WSN based fire alarm system using Arduino [8]. The system is purely based on sensor alerts and it lacks other features like camera, web server etc.

Kumar *et al.* in their paper described the surveillance technique using IP camera and Arduino board [9]. In this paper, user can view remote desktop using team viewer application whenever he needs to monitor his home from outside. This system is not sending any notification to user whenever any unusual event occurs in his home. User has to monitor his home continuously and also it lacks sensor based alerts.

Most of the previous papers on security systems are utilizing zigbee based WSN. But it has limited range and bandwidth. Some previous papers describe only sensor alert techniques and it lacks the video surveillance, web servers, live streaming etc. Almost all the systems are implemented using microcontroller or computer. Since microcontroller cannot perform multiple functions at a time, the same can be achieved by making use of a computer. But computer is very expensive and consumes more power.

In this paper, we introduce a newly revised security system using Raspberry Pi and NodeMCU (IoT/WiFi module) which integrates WSN with video surveillance techniques. Raspberry Pi is a low cost, low power, single board computer which can handle multiple functions like a normal computer. Intruder detection with face recognition feature and fire detection are the prime features of this system. This system is purely based on WiFi connectivity. Here, we use WiFi module for wireless transmission instead of zigbee. Having WiFi connectivity is an added advantage for any system, data can be fetched from anywhere and it can be moved to cloud for storage and monitoring. Also, it has long range and high bandwidth, so that it is perfect for streaming video, sending email etc. A normal WSN consists of one or more sensors, a microcontroller to process the sensor output and a transceiver module for wireless transmission of data. In our work, microcontroller and wireless transmitter in a normal WSN is replaced by a single IoT module. Also, the receiver section is connected to WiFi, which eliminates the need of separate wireless receiver module. Thus, the IoT module makes the system cost effective, compact, globally controllable and accessible. The objective of our paper is to design and implement a low cost, reliable, energy efficient, long range, globally accessible, storage effective surveillance system using Raspberry Pi SBC, GSM Modem and NodeMCU (IoT Module).

Organization of the paper is as follows: Section II describes the architecture of the system. In Section III, design and working of the entire system are discussed and in section IV, the implementation results are shared. In section V, the advantages of system are given. The conclusion is drawn in section VI.

## II. IoT BASED SYSTEM ARCHITECTURE

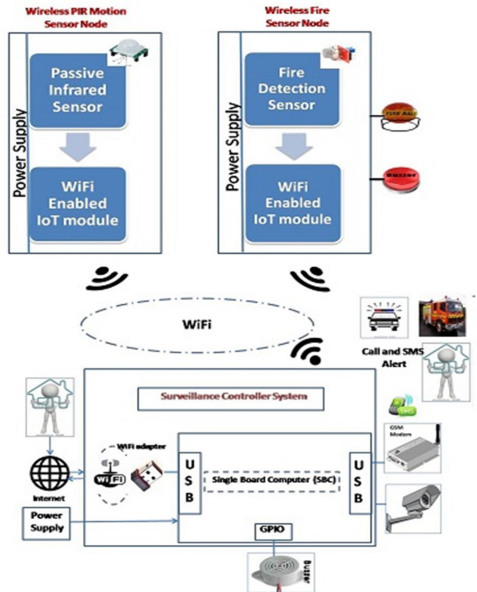

Fig. 2. IoT based System Architecture

The detailed architecture of IoT based surveillance system is shown in Fig. 2. The entire system is divided into two parts: *WiFi enabled sensor nodes and Surveillance controller system (master node)*. It is mandatory that both sensor and master node should be on same network. Being the sensor node as wireless, they can be easily installed anywhere inside home/building. Sensor nodes follow modular design so that any number of nodes can be added or removed.

### A. WiFi enabled sensor nodes

Here, sensor nodes are organized as modules. Each module consists of a sensor node followed by a NodeMCU. It is a low cost open source WiFi module used for IoT applications. The sensor is connected to the IoT module which will process the sensor output and update the sensor status to master node via WiFi.

There are two types of sensor nodes in this system. Passive Infrared (PIR) Motion sensor node and Fire detection sensor node. PIR node will detect the presence of the intruder and it can be installed in entry restricted areas or other critical areas. Fire node will detect the presence of fire, triggers buzzer and activates the fire safety device i.e.; a fire ball.

### B. Surveillance Controller System (Master Node)

It is built on Raspberry Pi (RPi) SBC with a Linux based operating system 'Raspbian' installed into it. RPi is a low cost, low power, credit card sized computer. Master node consists of Webcam for video surveillance, GSM for remote notification, and buzzer for emergency alert. This node handles various functions such as managing the sensor feeds, camera feeds, facial feature extraction and comparison, SMS and call alerts via GSM, email alerts, video streaming etc.

NodeMCU is programmed to work as a WiFi server. Here the system is connected to WiFi through a USB WiFi adapter. The master node acts according to the sensor updates from sensor node.

When master node receives the valid sensor status of PIR sensor, Webcam is activated and image of intruder is captured. From the captured frame, facial features are extracted and compared with those in the database. Once the face recognition is carried out, system checks for known and unknown people. If the person is unknown, then the user will be notified using email notification with the picture of unknown person as attachment. If the person is known also, the user will get a notification as email with the attachment of image of the scene with a comment "'Name X' entered in your home". Then, Webcam starts recording the event and the Webcam server places the live streaming of event into internet.

The system triggers buzzer for audio signaling and enables GSM modem to send SMS and call notification to the user upon detecting the event. User can watch the live video of the event happened in his home from anywhere using the IP address of RPi. If the video contains a known person, then user will send a reply SMS as 'Found OK' to GSM modem to stop the surveillance activities such as video recording, live streaming, buzzing etc. Else, the user will send a reply SMS as 'Inform Authorities' so that GSM modem would inform the concerned authorities such as police, fire force etc. Video of the event will be saved in RPi's SD card and it will be deleted once it is send to user by email. Thus it saves storage space in SD card. Also, the video files can be stored safely even if master node is seized or damaged. When fire is detected, the system only enables GSM modem to notify the user.

## III. DETAILED SYSTEM DESIGN

### A. Hardware Design

The system is designed as an embedded system which has two parts: WiFi enabled sensor nodes and Surveillance Controller System. The hardware design of each part is as follows:

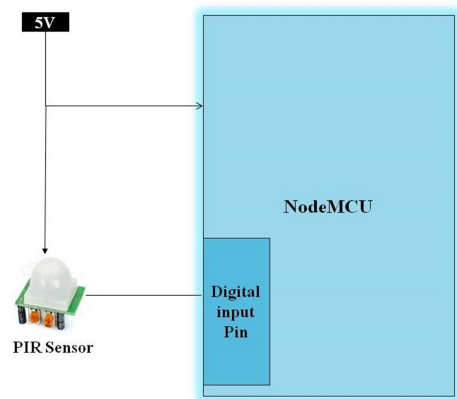

Fig. 3. PIR Sensor Node Schematic

*1) Hardware design of PIR sensor node*

Fig. 3 shows the schematic block diagram of the PIR sensor node. It consists of NodeMCU module for interfacing the sensor. NodeMCU is a low cost, low power, smart, open source WiFi module operates at a frequency of 80 MHz. It is embedded with Tensilica L106 32-bit micro controller (MCU) and is used for carrying out IoT applications. It has 13 GPIO pins, micro usb interface for power and flash and a PCB antenna with a range of upto 400 m. It is very easy to use since the size is very small.

PIR sensor (HC-SR501) is a three pin device which works from 5V. It detects motion by measuring the changes in infrared radiation level of human beings. The presence of intruder is indicated by a logic high value at the output pin. It receives the sensor output and displays the sensor status in a webpage. The webpage can be viewed in browser using the IP Address of NodeMCU. Master node can get the sensor values from webpage since both the sensor and master node are connected to same WiFi server.

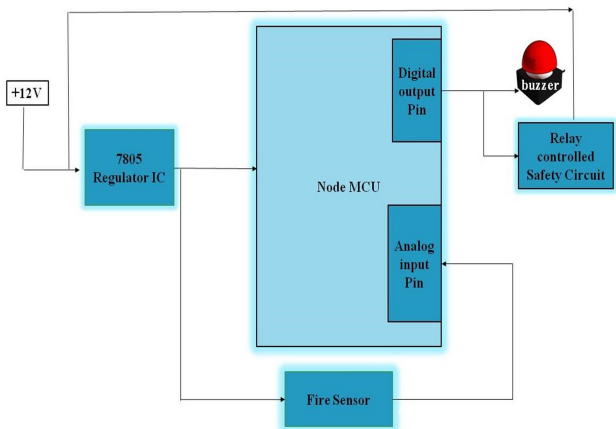

Fig. 4. Fire Sensor Node Schematic

*2) Hardware design of fire detection sensor node*

Fig. 4 shows the schematic block diagram of Fire sensor node. It consists of Temperature sensor LM35 for detecting fire (fire is indicated by an increase in temperature). The output of LM35 is analog in nature and it is connected to analog pin of NodeMCU. When the presence of fire is detected, buzzer buzzes and fire safety device is activated. The operation of fire safety device is controlled by a relay. We use 12V power supply since the selected safety device need 12V to work. Since NodeMCU will not support more than 5V, 12V supply is regulated to 5V and fed to NodeMCU. To control the operation of the safety device, 12V relay is used. Relay is driven by a transistor which is connected to the digital pin of MCU. The presence of fire will turn on the transistor relay driver circuit and safety device is activated.

*3) Hardware design of master node*

Fig. 5 shows the hardware design of master node: It consists of Raspberry Pi, USB Web Camera, GSM modem and a buzzer. RPi is connected to WiFi using WiFi adapter. USB camera is connected to RPi via USB port and buzzer is connected via GPIO pins. GSM modem is interfaced with RPi using serial port pins RX and TX. RPi is powered using a USB 5V supply and GSM modem is powered using a 12V adapter.

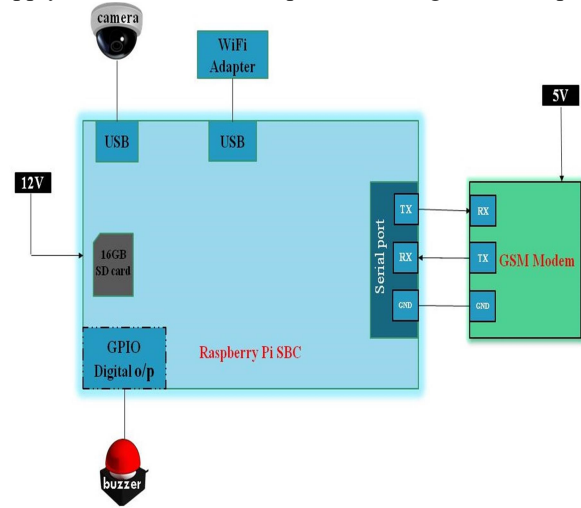

Fig. 5. Master Node Schematic

Fig. 5 shows the hardware design of master node: It consists of Raspberry Pi, USB Web Camera, GSM modem and a buzzer. RPi is connected to WiFi using WiFi adapter. USB camera is connected to RPi via USB port and buzzer is connected via GPIO pins. GSM modem is interfaced with RPi using serial port pins RX and TX. RPi is powered using a USB 5V supply and GSM modem is powered using a 12V adapter.

Camera is placed facing the main entrance door. It will be in inactive state normally and will move to active state when intruder's presence is detected by the PIR sensor. Then it will capture the frame, executes face recognition algorithm, alerts the user, records the event, places the live stream into internet, informs the authorities or ceases the activities based on the received SMS from user and stores the video files in SD card for future reference for a pre-defined time. User can watch the stored video in case he missed live stream.

*B. Software Design*

Fig. 6 shows the detailed software design of the system. It is divided into two sections. Wireless sensor node software design and master node software design.

*1) Wireless sensor node software design*

Sensor node is made as wireless by programming the NodeMCU board to work in WiFi. It is programmed with Arduino integrated Development Environment (IDE) using C language. We need to follow certain steps like installation of required libraries and packages, selection of programming board etc. in IDE to program NodeMCU. After flashing our code into NodeMCU, the board will be connected to WiFi and it will be turned into a WiFi web server. The IP address of the board can be seen in the serial monitor window. Web server can be accessed by typing the NodeMCU's IP address in web browser. Sensors connected to the NodeMCU will sense the event and the microcontroller will send the sensor status to webpage. Occurrence of an event is indicated in webpage by a logic high value.

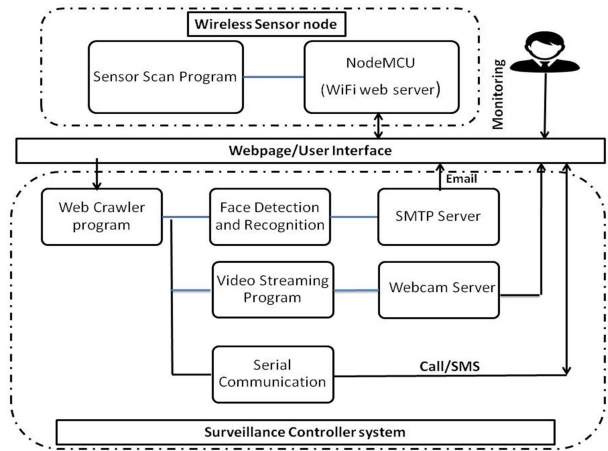

Fig. 6. Software Design of Proposed System

*2) Master node software design*

A Linux based operating system (OS) called Raspibian OS is installed into SD card. RPi's desktop is accessed from laptop using ssh remote login. Then all the required libraries and softwares such as Open Source Computer Vision (OpenCV), Python, streaming software (Motion), web server etc. were installed into it. We have used OpenCV-python for writing application programs.

Whenever there is an event in sensor node, master node will execute various functions. Sensor node will update its status regularly to a webpage. Master node will crawl the sensor node's webpage using a web crawler. If the crawled data has a logic high sensor value, then the master node will trigger a python script which executes a sequence of functions such as face detection and recognition, email alert, live streaming, SMS and call alert etc.

2a) *Face Detection and* Recognition

The system captures image from the camera installed, then face detection and recognition process is carried out. Detection is done using Haar feature-based cascade classifiers, which is an effective way of detecting face and is proposed by Paul Viola, and Michael Jones [10].

Haar cascade function is a cascade of positive and negative images. Positive images are faces and negative images are nonfaces. Facial Features are extracted from it using haar features which are shown in Fig. 7. Each feature value is the difference between sum of pixel values in white region to that of the black region. Haar features are in different size and shape, and we can get infinite number of facial features from it. We use different methods such adaboost and cascade of classifiers to simplify the summation and to filter out irrelevant features from it. A 24 × 24 window is taken and apply these features into it. Features are grouped under different stages. A window which passes through all the stages will be classified as a face [10].

Face recognition is followed after the detection process. Recognition usually requires two phases, training and testing phase. During training phase, we train our data base. Image of people are taken under various illuminations and lighting conditions with different expressions. During the testing phase, the image of the person to be tested is compared with those in the data base to find the match. Face recognition is carried out with Fisher Face algorithm [11] which is an advanced version of Principal Component Analysis (PCA) algorithm and it is shown in Fig. 8.

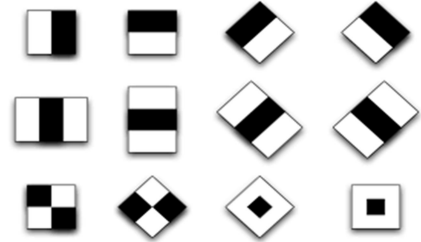

Fig. 7. Haar Features

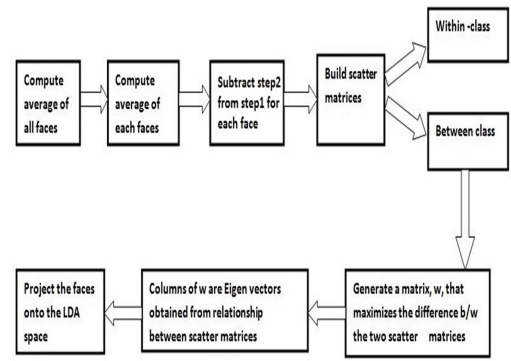

Fig. 8. Fisher Face Algorithm Flowchart

Each individual is grouped under a class and each of these classes has images taken under different illumination, lighting and with different expressions. Then, average of all the classes are computed, difference between them are taken, scatter matrixes are build accordingly within class as well as between class and variances are computed. A weight matrix is build which maximizes the difference between two scatter matrices. Face images are projected on to this Linear Discriminant Analysis (LDA) space.

*Fisher Face Algorithm*
The fisher face method used for face recognition uses both PCA and LDA for producing a projection matrix Fisher face method takes advantage of within class information, minimizing variation within each class and maximizing class separation. The image in the training set is represented by vector size (N × N) representing the size of the image. Using PCA, we can represent an image using M eigenvectors where M is the number of eigenvector used. The mathematical explanation of the face recognition is given below.
Take (N × M) image array and reshape into a ((N × M) × 1) vector. Then, calculate both the class mean $\mu_k$ and mean of all the samples $u$.

$$\mu_k = \frac{1}{N}\sum_{m=1}^{N^k} X_{K_m} \qquad (1)$$

$$\mu = \frac{1}{N_k} \sum_{K=1}^{N} x_k \qquad (2)$$

where, N - Total number of images
$x_k$ - Image from the database of the kth class
$N_k$ - Number of images in class k

$X_{k_m}$ - Image at index m of class k.

Now we have to determine both between class scatter matrix ($S_B$) and the within-class scatter matrix ($S_W$).

$$S_B = \sum_{K=1}^{C} N_K (\mu_K - \mu)(\mu_K - \mu)^T \quad (3)$$

$$S_W = \sum_{K=1}^{C} \sum_{x \in X_k}(x_k - \mu_k)(x_k - \mu_k)^T \quad (4)$$

where, C - Number of classes

The optimal Eigen vectors ($U_{opt}$) is found out by

$$U_{opt} = \arg\max \frac{|U^T S_B U|}{|U^T S_W U|} = \mu_1, \mu_2 \ldots \ldots, \mu_m \quad (5)$$

This equation can be simplified into a generalized eigen value equation:

$$S_B \mu_i = \lambda_i S_W \mu_i : i=1, 2\ldots\ldots m \quad (6)$$

The feature vector can be established using the equation as,

$$y_k = U^T x_k : k=1, 2\ldots\ldots m \quad (7)$$

Then project the faces to the LDA space for recognition results.

Both testing as well as training images are projected to this LDA space, test image values are compared with each of the images in the data base. A difference of the test image with each of the class is calculated, i.e. Euclidian distance. Test image is mapped to a particular class which has minimum Euclidian distance and which is less than a particular threshold value. If the Euclidian distance does not fall below the threshold to any of the class, then the image can be considered as unknown.

Once it shows a known/unknown person, the image of the respective person is send to the owner by email. In order to send email using python, we use Simple Mail Transfer Protocol (SMTP) object. SMTP object will access the SMTP server and the email is routed to the selected recipients.

2b) Live streaming

Live feeds from camera can be seen directly on a web browser by turning RPi into a Webcam server. It is done by executing the streaming software called motion which is capable of hosting live stream. We have to set various parameters such as Webcam streaming Port, frame rate etc. in the motion configuration file of RPi. Then, Webcam server will display the corresponding live video on a webpage. Video can be watched in local network by typing RPi's IP Address followed by the corresponding Port.

In order to watch live video from anywhere, we use Port Forwarding Technique. It is a technique where the internal IP Address and Port of a device is mapped to the external IP address and Port. Here, the IP Address and streaming Port of RPi is added in the Port mapping window of WiFi router. Thus, the RPi's IP Address and port is mapped to router's IP Address and port. Live video can be watched from anywhere using router's IP Address and Port. Video will be saved in RPi for future reference and it will be deleted once it is send to user by email.

2c) Serial Communication

RPi communicates with GSM modem via serial Port. RPi sends suitable Attention (AT) commands to GSM modem and it will respond back by sending/receiving SMS and making calls to the respective mobile numbers.

IV. RESULTS OF IMPLEMENTATION

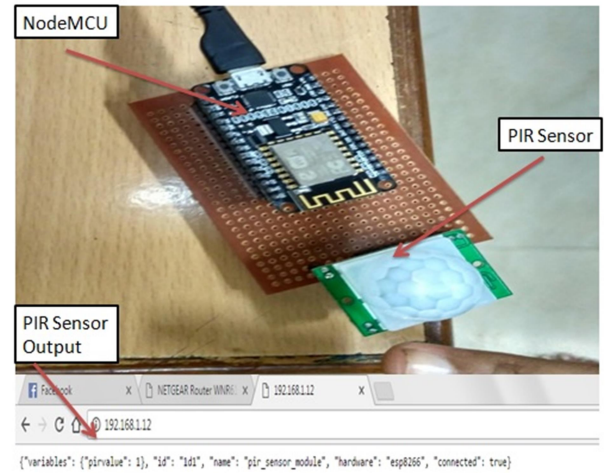

Fig. 9. Wireless PIR Sensor Node

Fig. 9 shows the PIR sensor node implemented using IoT module (NodeMCU). The output obtained from the PIR sensor is shown on a webpage. The IoT Module is powered using a normal USB 5V supply. The sensor gets 5V supply from MCU and it gives a logical output to the digital pin of MCU. In the presence of human being, it gives logic high at its output pin and it is indicated in webpage shown in Fig. 9.

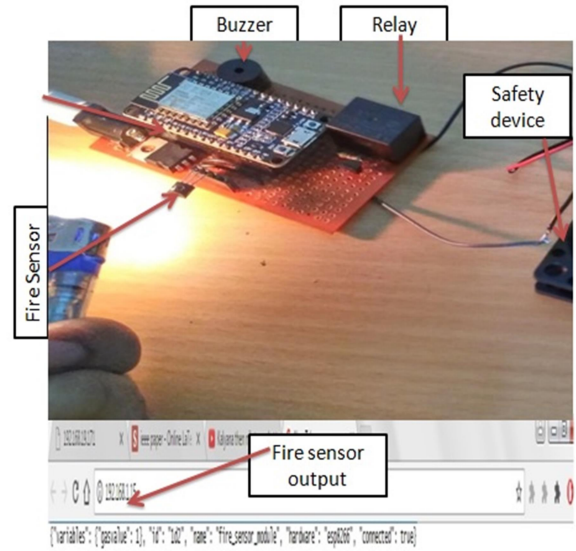

Fig. 10 Wireless Fire Sensor Node

Fig. 10 shows the fire sensor node developed using IoT module. The sensor status of fire sensor node is shown on the corresponding webpage. The output of sensor is connected to analog pin of NodeMCU. NodeMCU reads the analog output of sensor and makes a digital pin high when sensor senses fire. Buzzer and transistor relay driver circuit are connected to that digital pin. Thus buzzer buzzes and safety device gets activated when sensor detects fire.

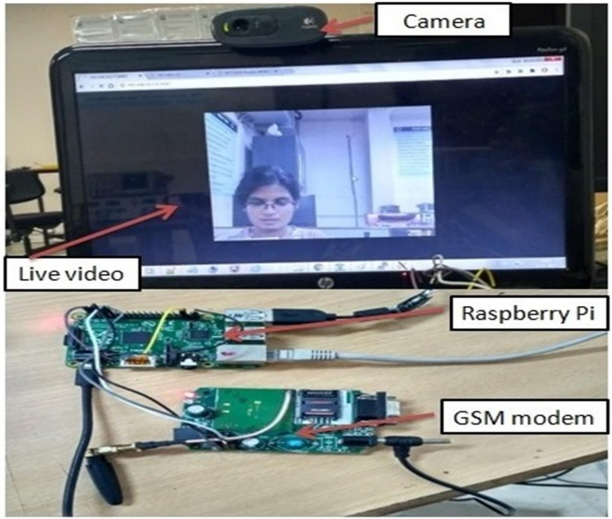

Fig. 11.  Surveillance Controller System

Fig. 11 shows the surveillance controller system implemented using Raspberry Pi. It includes buzzer, GSM Modem and Webcam. Raspberry Pi is powered by 5V adapter and GSM is powered by 12V adapter. The desktop of Raspberry Pi is accessed remotely from putty software which is installed on Laptop.

The screenshot of the live video is also shown in Fig. 11. User seeks surveillance on webpage using WiFi router's IP Address. This system is useful for the owner to get a remote view of his home and to keep an eye on his valuables.

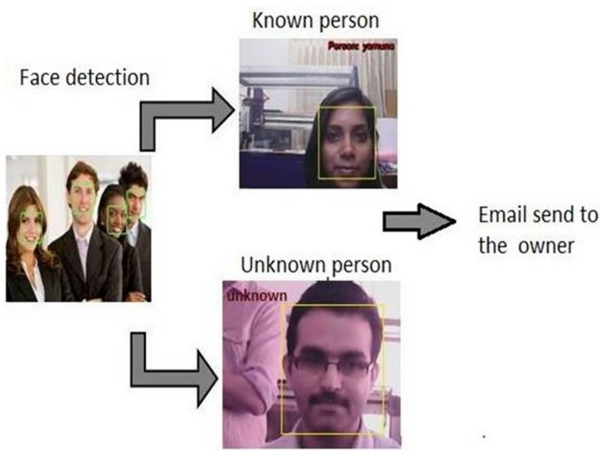

Fig. 12.  Face Detection and Recognition Results

Fig. 12 shows the implementation results of face recognition. Once the face recognition is carried out, system checks for known and unknown people. If there is a match, it will show the known person's name else, it will show unknown.

A provision of sending an email to the owner is provided. In the email, the image of the known/unknown person is also uploaded as proof.

Fig. 13.  Python Shell

Fig. 13 shows the sensor status obtained at the master node section. The sensor output crawled from webpage is shown in python shell. If the PIR sensor data crawled from webpage shows a logic high value, then camera is activated, image is captured, face detection and recognition is done and the image of known/unknown person is send to the user. After that motion software is triggered, video is recorded and notification is send to user via AT commands. Then required action is taken according to the responses from the user.

If the fire sensor data crawled from the webpage gives a logic high value, then GSM is activated and SMS and call notification is send to user.

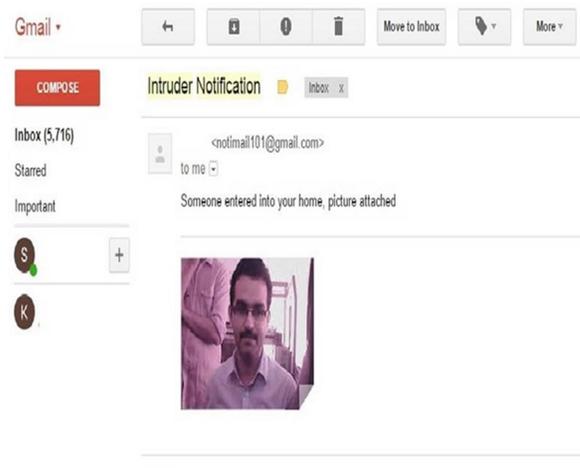

Fig. 14.  Screenshot of Email

Fig. 14 shows the screenshot of received Email with the picture of intruder as attachment.

## V.  ADVANTAGES OF SYSTEM

*Low cost:* The use of IoT module reduces the cost of wireless sensor nodes. In normal case, WSN will have a sensor, a microcontroller and a wireless transmitter module. At the receiver section, a wireless receiver module will be there. The function of microcontroller and a wireless transmitter module can be achieved by making use of a single IoT module. Since the receiver is connected to WiFi, wireless receiver module is not needed. Thus many nodes can be added and the cost of the entire system can be reduced.

*Compact*: Since size of IoT module is very small, it is very easy to handle and thus it makes entire system compact.

*Long Range*: IoT module gives extremely good working range. So for long range communication, we prefer this module. We are able to communicate with the module up to a distance of 400m.

*Worldwide monitoring and control:* The use of WiFi in embedded system makes the system globally controllable and accessible.

## VI. CONCLUSION

In this paper, we have designed and developed a real time surveillance system using IoT module and Raspberry Pi. It is an active surveillance system which will alert the user when the event happens. Intruder detection with face recognition and fire detection are the prime features of the system. Even though face recognition was a challenging task, we were able to identify the person with better accuracy. Live video streaming is an additional advantage of the system. We have created web servers which help the user to view the sensor status and the live video. This system also sends intruder's picture and the captured video to the owner by email. The use of NodeMCU makes the system cost effective, portable and compact. Most of the existing surveillance systems are costly and common people may not spend a lot for such systems. This system is designed with an aim that it can be used for all kind of people since security of every one's home should not be left behind.

## VII. FUTURE SCOPE

We have implemented only two sensor nodes in our project. More number of sensor modules like gas, pressure, humidity, pollution, etc. can be added to improve the efficiency of the system. We can use multiple cameras to develop a distributed surveillance system for monitoring multiple scenes. It will provide a more accurate model of the monitored scenes with multiple views. In order to improve the accuracy of face recognition, we have to use more advanced face recognition algorithm.

The risks to users of wireless technology have increased as the service has turned out to be more popular. So, we have to add proper wireless encryption techniques to prevent the unauthorized access. Also, we can apply several encryption algorithms to secure video transmission. We can create attractive webpage with proper authentication to watch the live stream. Also, we can provide button controls in webpage to control the working of sensor nodes.

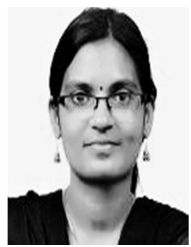 **Sruthy. S** received the B.Tech degree in Electronics and Communication Engineering from Calicut University, Kerala, India in 2012. She will be receiving her M.Tech degree in Electronics Design and Technology from National Institute of Technology, Calicut, India, in 2017.

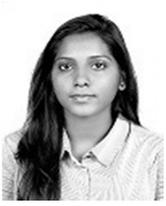
**S. Yamuna** received the B.Tech degree in Electronics and Communication Engineering from Calicut University, Kerala, India in 2014. She will be receiving her M.Tech degree in Electronics Design and Technology from National Institute of Technology, Calicut, India, in 2017.

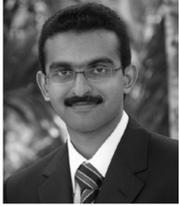
**Sudhish N. George** received the B.Tech degree in Electronics and Communication Engineering from M.G University, Kerala, India, in 2004, M.Tech degree in signal processing from Kerala University, India, in 2007 and PhD in multimedia security from National Institute of Technology Calicut, India, 2014. He is working as Assistant Professor in Department of Electronics and Communication, National Institute of Technology Calicut, India from 2010 onwards. His research interest is related to sparse signal processing and computer vision.